\documentclass[lettersize,journal, onecolumn]{IEEEtran}
\usepackage{amsmath,amsfonts}
\usepackage{algorithmic}
\usepackage{algorithm}
\usepackage{array}
\usepackage[caption=false,font=normalsize,labelfont=sf,textfont=sf]{subfig}
\usepackage{textcomp}
\usepackage{stfloats}
\usepackage{url}
\usepackage{verbatim}
\usepackage{graphicx}
\usepackage{cite}
\usepackage{physics}
\usepackage{amssymb, amsmath,amsfonts,bm,bbm,amsthm}
\usepackage{mathtools}
\allowdisplaybreaks
\usepackage[brazilian, capitalise]{cleveref}
\usepackage{pifont}

\newtheorem{theorem}{Theorem}

\newtheorem{proposition}[theorem]{Proposition}

\DeclareMathAlphabet\mathbfcal{OMS}{cmsy}{b}{n}

\newcommand{\Hilb}{\ensuremath{\mathcal{H}}}

\newcommand{\opname}[1]{\ensuremath{\operatorname{#1}}}

\renewcommand{\vu}[1]{\ensuremath{\hat{#1}}}
\renewcommand{\tr}{\ensuremath{\operatorname{tr}}}

\newcommand{\m}{\bar{m}}

\newcommand{\opa}{\vu{a}}
\newcommand{\opad}{\vu{a}^{\dagger}}

\newcommand{\opb}{\vu{b}}
\newcommand{\opbd}{\vu{b}^{\dagger}}

\usepackage{widetext}
\begin{document}

\title{Convergence of Density Operators and Security of Discrete Modulated CVQKD Protocols}

\author{Micael Andrade Dias and Francisco Marcos de Assis
\thanks{This paper was produced by the IEEE Publication Technology Group. They are in Piscataway, NJ.}
\thanks{Manuscript received April 19, 2021; revised August 16, 2021.}}



\maketitle

\begin{abstract}
  This communication deals with the problem of bounding the approximation error on weak convergence of mixed coherent state towards a Gaussian thermal state. In the context of CVQKD with discrete modulation, we develop expressions for two specific cases. The first one is the distance between the Gaussian equivalent bipartite state and a reference Gaussian modulated (GG02) and the second one is for the trace distance between the constellation and a thermal state with same photon number. Since, in the convex set of density operators, weak convergence implies convergence in the trace norm, knowing how fast the sequence gets close to the equivalent Gaussian state has implication on the security of QKD Protocols. Here we derive two bounds on the $L_1$ distance, one of them related with an energy test that can be used in the security proof.
\end{abstract}



\section{Preliminaries}

In this section, we discuss the discrete modulated (DM) CVQKD protocol and the framework of our analysis. Consider a complex-valued random variable $X_n$, $n\geq1$ with alphabet $\mathcal{X}_n$ and $Pr[X_n = x] = p_{X_n}(x)$. For simplicity and compatibility with usual digital communication systems, we assume that $|\mathcal{X}_n| = (n+1)^2=m^2$ and whenever it is clear from the context, we drop the subscript and use only $p(x_i)$. We also use the notation $[N] = \qty{1,2,\cdots,N}$ and assume that $X_n$ is symmetric around the origin\footnote{This assumptions comes without loss of generality as the constellations approaching the AWGN channel capacity are usually symmetric.}, that is, $p(x_i) = p(-x_i)$. The $m^2$-state DM-CVQKD prepare and measure protocol induced by $X_n$ works as follows.

\begin{enumerate}
	\item{\textit{State Preparation -} At each round, Alice draws $x$ from $X_n$ and prepares a coherent state $\ket{x}$. The modulation scheme is represented by the ensemble $\mathcal{A} = \qty{\ket{x_k}, p(x_k)}_{k=1}^{m^2}$, $x_k\in\mathcal{X}_n$, which is the mixture $\vu\rho_{X_n} = \sum_{x\in\mathcal{X}}p(x)\op{x}$ and the register $\bm{X}'$ stores Alice's random experiment outcomes.}
	
	\item{\textit{Quantum Transmission and Measurement -} The prepared state is sent through a one-mode quantum channel $\mathcal{N}_{A\rightarrow B}$ such that Bob observes the mixture $\vu\rho_B = \mathcal{N}_{A\rightarrow B}(\vu\rho_{X_n})$ and performs heterodyne detection, denoted by the operator $\mathcal{M}_{B\rightarrow Y}$. Measurement results are stored in $\bm{Y}' = \mathcal{M}_{B\rightarrow Y}(\vu\rho_B)$.}
	
	\item{\textit{Sifting - }After the conclusion of $N$ rounds, Alice and Bob agree on a small random subset of $I_{test}\subset[N]$ to form the test set to be used in parameter estimation. The values of $\bm{X}'$ and $\bm{Y}'$ indexed by $I_{test}$ are publicly announced and then discarded. The remaining raw key values are indexed by $I_{key}=[N]\setminus I_{test}$ and represented by $\bm{X}$ and $\bm{Y}$ on Alice and Bob's sides, respectively.}
	
	\item{\textit{Parameter estimation - } Once the test set is defined, Alice and Bob use them to estimate the quantum channel parameters, mainly by estimating the data first and second moments. Based on the estimated values, they evaluate whether it is possible to distill a secret key and if it is not, they abort the protocol and go back to step (i).}
	
	\item{\textit{Data Estimation - } Before information reconciliation, Bob performs an estimation procedure in order to retrieve a sequence $\hat{\bm{X}} = \theta(\bm{Y})$, where $\theta$ is some estimation function. If $X_n$ has uniform distribution, i.e., $p(x_i) = \frac{1}{n}$, $\theta$ is a minimum distance estimator
	\begin{equation}
		\hat{\bm{X}}[j] = \arg\min_{x\in\mathcal{X}_n}||\bm{Y}[j] - x||^2,
	\end{equation}
	\noindent where $\hat{\bm{X}}[j]$ (respec. $\bm{Y}[j]$) denotes the $j$-th element of $\hat{\bm{X}}$ (respec. $\bm{Y}$). In the case of nonuniform distribution, which will be the relevant case, $\theta$ is the \textit{maximum a posteriori} (MAP) estimator,
	\begin{equation}
		\hat{\bm{X}}[j] = \arg\max_{x\in\mathcal{X}_n}p_{X_n|Y}(x|\bm{Y}[j]),
	\end{equation}}
	
	\item{\textit{Reconciliation and Privacy Amplification - } Alice and Bob choose a suitable error correction protocol and privacy amplification, which must agree with the quantities evaluated during parameter estimation. They then apply error correction and privacy amplification to generate the secret key.}
\end{enumerate}
\section{On the Convergence of Constellations}

\begin{theorem}[Operator norm convergence \cite{holevo2019}]\label{th:weak-norm-convergence}
    Let $\qty{\vu\sigma_n}$ be a sequence of density operators in $\mathcal{D}(\Hilb)$ weakly converging to $\vu\sigma$. Then $\qty{\vu\sigma_n}$ converges in the norm to $\vu\sigma$.
\end{theorem}

A direct consequence of \Cref{th:weak-norm-convergence} is that the density operators representing constellations of coherent states also converge in trace norm to the Gaussian mixture of coherent states, $\vu\rho_{X_G}$. However, it is of interest to know the convergence speed of the constellation. The next proposition provides an upper bound for the distance between $\vu\rho_{X_n}$ and $\vu\rho_{X_G}$ by considering a reduction in the relevant subspace dimension.

\begin{proposition}[Approximation error of convergent sequences]\label{prop:approximation-error}
Let $\qty{\vu\rho_{X_n}}$ be a sequence of density operators weakly converging to $\vu\rho_{X_G}$ and let $P_d = \sum_{k=0}^{d-1}\op{k}$. Then, $\norm{ \vu\rho_{X_n} - \vu\rho_{X_G} } = O((\frac{\bar{m}}{\bar{m}+1})^d)$.
\end{proposition}

\begin{proof}		
Define the projectors $P_d = \sum_{k=0}^{d-1}\op{k}$ and $Q = I - P_d$ such that 
\begin{equation}
\varepsilon = \tr(\vu\rho_{X_G}Q) = \sum_{k=d}^\infty\frac{\bar{m}^k}{(\bar{m}+1)^{k+1}} = \qty(\frac{\bar{m}}{\bar{m}+1})^d.
\end{equation}
Then, for any operator $A\in\mathcal{B}(\Hilb)$, 
\begin{equation}\label{eq:ineq-epsilon}
|\tr(\vu\rho_{X_G}QA)|\leq\norm{\vu\rho_{X_G}Q}_1\norm{A} = \varepsilon\norm{A}.
\end{equation}
Thus, following the development in \cite[Theorem 3]{wehrl1976} and using inequality \eqref{eq:ineq-epsilon}, we have $|\tr[(\vu\rho_n - \vu\rho)A]| \leq 6\cdot\varepsilon\norm{A}$, which yields
\begin{align}
\norm{\vu\rho_n - \vu\rho}_1 &= \sup_{\norm{A}\leq1}|\tr[(\vu\rho_n - \vu\rho)A]| \\
                            &= \sup_{\norm{A}\leq1}6\cdot\varepsilon\norm{A} \\
                            &\leq 6\qty(\frac{\bar{m}}{\bar{m}+1})^d.
\end{align}
\end{proof}

The limit provided by \Cref{prop:approximation-error} is related to the results developed in \cite{leverrier2013}, where the unconditional security of the Gaussian modulation protocol against arbitrary attacks is reduced to collective attacks by using an energy test that truncates the Fock space. The next proposition relates the convergence of the sequence of operators to the convergence of the eigenvalues and eigenvectors. To address this issue, we will use some results from the theory of perturbation of linear operators.

The following theorem provides an upper bound for the distance between the eigenvalues of linear operators, which will be useful in analyzing the convergence of the eigenvalues of the convergent sequence of density operators \cite[Chapter V, Theorem 4.10]{kato1995}.

\begin{theorem}[Perturbation of eigenvalues]\label{th:spectral-distance}
Let $T$ be a self-adjoint operator, and $A\in\mathcal{B}(\Hilb)$ symmetric. Then, $T_A=T+A$ is self-adjoint, and $\opname{dist}(\Lambda(S), \Lambda(T_A))\leq\norm{A}_1$, where $\Lambda(\cdot)$ denotes the eigenvalue spectrum of the operator.
\end{theorem}

\begin{proposition}\label{prop:spectral-convergence}
Consider the convergence $\vu\rho_{X_n}\rightarrow\vu\rho_{X_G}$ and the spectral decompositions $\vu\rho_{X_n} = \sum_{k=1}^N\lambda_{k,n}\op{\phi_{k,n}}$ and $\rho_{X_G} = \sum_{n=0}^\infty \frac{\m^n}{(\m+1)^{n+1}}\op{n} = \sum_{n=0}^\infty \lambda_n\op{n}$. Then,
\begin{enumerate}
\item For all $k\in\mathbb{N}$, $\lim_{n\rightarrow\infty}|\lambda_{k,n} - \lambda_k| = 0$, 
\item For all $k\in\mathbb{N}$, $\lim_{n\rightarrow\infty}\norm{\op{\phi_{k,n}} - \op{k}}_1 = 0$.
\end{enumerate}
\end{proposition}

\begin{proof}
Take $T_A = \vu\rho_{X_n}$ and $T = \vu\rho_{X_G}$ so that $A = \vu\rho_{X_n} - \vu\rho_{X_G}$. For the first part of the proposition, \Cref{th:spectral-distance} results in $\opname{dist}(\Lambda(\vu\rho_{X_n}), \Lambda(\rho_{X_G})) \leq \norm{\vu\rho_{X_G} - \vu\rho_{X_n}}$, which can be made arbitrarily small with an increasing $n$ since the weak convergence of operators in $\mathcal{D}(\Hilb)$ coincides with the trace norm convergence. For the second part, take an eigenvalue $\lambda_k$ of $T$ and a smooth closed curve $\Gamma_{\lambda_k}$ in $P(T_A)$ containing $\lambda_k$ and no other eigenvalues of $T$. Using the expression in \eqref{eq:projector},
\begin{align}
P_A &= -\frac{1}{2\pi i}\int_{\Gamma_{\lambda_k}}\qty[R(z) + R(z)\sum_{p=1}^\infty\qty[-A\cdot R(z)]^p]\dd[2]{z},\\
&= P_{\lambda_k} - \frac{1}{2\pi i}\int_{\Gamma_{\lambda_k}}R(z)\sum_{p=1}^\infty\qty[-A\cdot R(z)]^p\dd[2]{z},
\end{align}
where $P_{\lambda_k}$ is the projector corresponding to the eigenvalue $\lambda_k$ of $T$, and the remaining term is related to the \textit{perturbation} $A$. Expanding the sum, we have

\begin{widetext}
    \begin{align}
        \sum_{p=1}^\infty\qty[-A\cdot R(z)]^p &= -AR(z) + AR(z)AR(z) - AR(z)AR(z)AR(z)+\cdots\\
    &= A\qty[-R(z) + R(z)AR(z) - R(z)AR(z)AR(z)+\cdots],
    \end{align}
    resulting in
    \begin{equation}
        P_A = P_{\lambda_k} - \frac{1}{2\pi i}\int_{\Gamma_{\lambda_k}}R(z)A\qty[-R(z) + R(z)AR(z) - R(z)AR(z)AR(z)+\cdots]\dd[2]{z}.
    \end{equation}
    Then, the difference between the projectors is
    \begin{align}
        \norm{P_{\lambda_k} - P_A} &= \frac{1}{2\pi}\norm{ \int_{\Gamma_{\lambda_k}}R(z)A\qty[-R(z) + R(z)AR(z) \cdots]\dd[2]{z} }
        \\
        &\leq \frac{1}{2\pi}\int_{\Gamma_{\lambda_k}}\norm{ R(z)A}\norm{\qty[-R(z) + R(z)AR(z) \cdots]}\dd[2]{z}\\
        &\leq \frac{1}{2\pi}\int_{\Gamma_{\lambda_k}}\norm{R(z)}\cdot\norm{A}\cdot\norm{-R(z) + R(z)AR(z) \cdots}\dd[2]{z}\\
        &= \frac{\norm{A}}{2\pi}\int_{\Gamma_{\lambda_k}}\norm{R(z)}\cdot\norm{-R(z) + R(z)AR(z) \cdots}\dd[2]{z},
    \end{align}
\end{widetext}

which can be made arbitrarily small since $\norm{A} = \norm{\vu\rho_{X_G} - \vu\rho_{X_n}}$.
\end{proof}

\begin{proposition}\label{prop:bipartite-state-convergence}
    Let $X_G\sim\mathbb{C}\mathcal{N}(0,\m)$ and $\qty{X_n}_{n\in\mathbb{N}}$ be a sequence of random variables converging in distribution to $X_G$, and let $\ket*{\Phi_{AB_n}} = (\mathbbm{1}_A\otimes\vu\rho_{X_n}^{\frac12})\sum_{n=0}^\infty\ket{n}\ket{n}$ be a purification of $\vu\rho_{X_n}$. Then $\vu\rho_{AB_n}\rightarrow\vu\rho_{AB}$, where $\vu\rho_{AB_n} = \op*{\Phi_{AB_n}}$ is the purification of the constellation, and $\vu\rho_{AB} = \op\nu$ is the EPR state with $\nu = 2\m+1$.
    \end{proposition}
    
    \begin{proof}
    \begin{align}
    \vu\rho_{AB_n} &= \op*{\Phi_{AB_n}}\\ 
    &= \qty(\sum_l\ket{l}\otimes\vu\rho_{X_n}^{\frac12}\ket{l})\qty(\sum_m\bra{m}\otimes\bra{m}\vu\rho_{X_n}^{\frac12})\\
    &= \sum_{l,m}\op{l}{m}\otimes\vu\rho_{X_n}^{\frac12}\op{l}{m}\vu\rho_{X_n}^{\frac12}\\\nonumber
    &= \sum_{l,m}\op{l}{m}\otimes\sum_{j,k}\sqrt{\lambda_{j,n}\lambda_{k,n}}\op{\phi_{j,n}}\op{l}{m}\op{\phi_{k,n}}\\
    &\overset{(n)}{\rightarrow} \sum_{l,m}\op{l}{m}\otimes\sum_{j,k}\sqrt{\lambda_j\lambda_k}\braket{j}{l}\braket{m}{k}\op{j}{k}\\
    &= \sum_{l,m}\op{l}{m}\otimes\op{l}{m}\cdot\frac{1}{\m+1}\qty[\qty(\frac{\m}{\m+1})^{\frac12}]^{l+m},
    \end{align}
    where the convergence of eigenvalues and eigenvectors is according to \Cref{prop:spectrum-convergence}. Taking $\m = \sinh^2(r)$ and $\lambda = \tanh(r)$, we have $1/(\m+1) = 1-\tanh^2(r) = 1 - \lambda^2$ and $\sqrt{\m/(\m+1)} = \sqrt{\tanh^2(r)} = \sqrt{\tanh^2(-r)} = \tanh(-r) = -\tanh(r) = -\lambda$. Thus,
    \begin{equation}
    \vu\rho_{AB_n} \rightarrow \sum_{l,m}\op{l}{m}\otimes\op{l}{m}\cdot(1-\lambda^2)\cdot(-\lambda)^{l+m} = \vu\rho_{AB}.
    \end{equation}
    
    Alternatively, consider $\ket{\psi},\ket{\gamma}\in\Hilb_A\otimes\Hilb_B$ arbitrary such that, without loss of generality, they can be decomposed as $\ket{\psi} = \ket{\psi}_A\ket{\psi}_B$ and $\ket{\gamma} = \ket{\gamma}_A\ket{\gamma}_B$. Then,
    \begin{align}
    \mel{\psi}{\vu\rho_{AB_n}}{\gamma} &= \bra{\psi}_A\bra{\psi}_B\qty(\sum_{j,k=1}^{(n+1)^2}\sqrt{\lambda_j\lambda_k} \ket{\phi_j}\ket{\phi_j}\bra{\phi_k}\ket{\phi_k} )\ket{\gamma}_A\ket{\gamma}_B\\
    &= \sum_{j,k=1}^{(n+1)^2}\sqrt{\lambda_j\lambda_k}\prescript{}{A}{\braket{\psi}{\phi_j}}\prescript{}{B}{\braket{\psi}{\phi_j}}\braket{\phi_k}{\gamma}_A\braket{\phi_k}{\gamma}_B\\
    &\overset{n}{\rightarrow} \sum_{j,k=1}^{\infty}\frac{\m}{\m+1}\qty[\qty(\frac{\m}{\m+1})^\frac12]^{j+k}\prescript{}{A}{\braket{\psi}{j}}\prescript{}{B}{\braket{\psi}{j}}\braket{k}{\gamma}_A\braket{k}{\gamma}_B\\
    &= \mel{\psi}{\vu\rho_{AB}}{\gamma},
    \end{align}
    so that $\vu\rho_{AB_n}$ converges weakly to $\vu\rho_{AB}$ and, consequently, converges in norm.
    \end{proof}
    
    Having dealt with the convergence of the purification of the bipartite state that purifies the constellation, the next step is to address the covariance matrix, specifically the lower bound $Z^*$ for the covariance between different-mode quadratures. As discussed earlier, the amount of information available to the eavesdropper is inversely proportional to the magnitude of $Z^*$, and the results previously discussed show the secret key rate of QAM-like modulations close to the rates for continuous Gaussian modulation indicate that $Z^*$ approaches $Z$.
    
    Among the relevant terms, $V_A$ and $V_B$ depend only on the average energy of the modulation scheme, which is a protocol parameter and does not depend on the specific type of constellation shape. Therefore, the secret key rate, being a function of the covariance matrix, where they differ only in the off-diagonal term (from the perspective of a block matrix), leads to using the Hilbert-Schmidt inner product for matrices $\ev{A,B} = \tr(A^\dagger B)$ and the induced norm. The Hilbert-Schmidt distance $d_{HS}(\cdot,\cdot)$ between the matrices
    \begin{align}
    \bm\Gamma(\vu\rho_{AB}) = \mqty(V_A\bm{I} & Z\bm\sigma_z \\ Z\bm\sigma_z & V_B\bm{I})  & & \bm\Gamma(\vu\rho_{AB_n}) = \mqty(V_A\bm{I} & Z_n^*\bm\sigma_z \\ Z_n^*\bm\sigma_z & V_B\bm{I})
    \end{align}
    will be
    \begin{widetext}
        \begin{align}
            d_{HS}(\bm\Gamma(\vu\rho_{AB}), \bm\Gamma(\vu\rho_{AB_n})) &= \qty[\ev{\bm\Gamma(\vu\rho_{AB}) - \bm\Gamma(\vu\rho_{AB_n}), \bm\Gamma(\vu\rho_{AB}) - \bm\Gamma(\vu\rho_{AB_n})}]^\frac12\\
            &= \qty[\tr(\bm\Gamma(\vu\rho_{AB}) - \bm\Gamma(\vu\rho_{AB_n}))^2]^\frac12,\\
            &= \qty(\tr\qty[\mqty(\bm{0} & (Z - Z_n^*)\bm{I} \\ (Z - Z_n^*)\bm{I} & \bm{0} )^2])^\frac12 \\
            &= \sqrt{4(Z - Z^*)^2}\\
            &= 2|Z - Z_n^*|\\
            &= 2|2\sqrt{\tau}\sqrt{\m^2+\m} - \sqrt{\tau}\ev{\opa\opb+\opad\opbd}{\Phi_{AB_n}} + \sqrt{2\tau\xi w}|\\
            &= 2|2\sqrt{\tau}\sqrt{\m^2+\m} - \sqrt{\tau}\tr[\vu\rho_{AB_n}(\opa\opb+\opad\opbd)] + \sqrt{2\tau\xi w}|\\
            &\rightarrow 0,
        \end{align}
    \end{widetext}
    since $w$ depends on $\vu\rho_{X_n}$ and, in the limit $\lim_{n\rightarrow\infty}\vu\rho_{X_n} = \vu\rho_{X_G}$, $w\rightarrow0$; similarly, due to the convergence of the bipartite state, $\tr[\vu\rho_{AB_n}(\opa\opb+\opad\opbd)] \rightarrow 2\sqrt{\m^2+\m}$.
\section{Unconditional security}
The two techniques used to reduce the class of arbitrary attacks to collective attacks are effective in the context of QKD protocols with discrete variables, such as BB84, but they are not directly applicable in the analysis of protocols with continuous variables due to the dimension of the Hilbert space. One solution is to use an energy test before the protocol's execution in conjunction with a projector in a finite-dimensional subspace $d$. The idea is that if the states measured by Bob have an average energy lower than a pre-established parameter, the probability of the bipartite state shared by Alice and Bob "residing" in a finite-dimensional subspace increases exponentially with $d$".

In particular, the security proof against arbitrary attacks developed in \cite{leverrier2013} for the protocol\footnote{The analyzed protocol uses Gaussian modulation combined with heterodyne detection.} $\mathcal{E}_0$ uses an energy test $\mathcal{T}$ and the projection $\mathcal{P}$ in a $d$-dimensional subspace, such that
\begin{equation}\label{eq:bound-general-attacks1}
    \norm{\mathcal{E} - \mathcal{F}}_\diamondsuit \leq \norm*{\tilde{\mathcal{E}} - \tilde{\mathcal{F}}}_\diamondsuit + 2\norm{(\mathbbm{1} - \mathcal{P})\circ\mathcal{T}}_\diamondsuit,
\end{equation}
where $\mathcal{E} = \mathcal{E}_0\circ\mathcal{T}$, $\mathcal{F} = \mathcal{S}\circ\mathcal{E}$, $\tilde{\mathcal{E}} = \mathcal{E}_0\circ\mathcal{P}\circ\mathcal{T}$, and $\tilde{\mathcal{F}} = \mathcal{S}\circ\mathcal{E}_0$. By employing the projector map in a finite-dimensional subspace, it was possible to use the results of the Postselection theorem to bound the first term on the right side of inequality \eqref{eq:bound-general-attacks1}, while the second term was upper-bounded by fixing the state of the diamond distance optimization to a product state, equivalent to the assumption of a collective attack, such that
\begin{equation}
    \norm{(\mathbbm{1} - \mathcal{P})\circ\mathcal{T}}_\diamondsuit \leq \epsilon_{test} = O\qty(\qty(\frac{\bar{m}}{\bar{m}+1})^d).
\end{equation}

The term $\epsilon_{\text{test}}$ precisely denotes the probability of the state shared by Alice and Bob passing the energy test $\mathcal{T}$ and being outside the relevant $d$-dimensional subspace for security analysis, which decreases exponentially with $d$ and is proportional to the distance between a thermal state and a convergent mixture of coherent states (Proposition \ref{prop:erro-aproximacao}). Therefore, it is expected that the unconditional security factor for a protocol with discrete modulation, where the constellation converges in distribution to a Gaussian curve, is proportional to the security developed in \cite{leverrier2013} for a protocol with continuous modulation.

However, there are still operational factors that may make the unconditional security proofs developed for Gaussian protocols incompatible with non-Gaussian modulation protocols, mainly due to the symmetrization procedures used to guarantee the necessary symmetry arguments. The Finetti and Postselection theorems assume that the protocol is invariant under permutations, and in \cite{leverrier2013,leverrier2015,leverrier2017}, the security (and composability) proofs of protocols with Gaussian modulation use passive linear operations (beam splitters and phase displacements) before measuring the states received by Bob to apply rotations in the phase space and exploit the protocol's invariance to rotations. As discussed in Section \ref{sec:problemas-formatacao}, the architecture of a transmitter for constellations with probabilistic shaping would be incompatible with random permutations of subsystems, preventing the protocol from having the required property in the assumptions of the security proof.

Regarding rotations in the phase space, a potential problem is that these rotations result in rotations in the vector representing the measurement outcomes. The possibility of reversing the effect of the symmetrization procedure by applying the inverse operation $R^{-1}$ does not solve the problem since DM-CVQKD protocols must use an optimized detection architecture that is tailored to the constellation format. Therefore, it is naive to assume that applying the symmetrization procedure to the multimode state and then reversing the effect by applying a rotation to the measurement data maintains the protocol's performance intact. To achieve this, it is necessary to ensure that the operation that performs rotations in the phase space commutes with the coherent pulse detection system and the constellation symbol estimation, whose description in terms of POVMs is not yet known \cite{chen2023}. Additionally, the symmetrization procedure, which corresponds to a network of beam splitters and phase displacements, becomes impractical in realistic scenarios with a large number of transmitted states.

Furthermore, it is important to emphasize that the assumption used in \cite{roumestan2021}, that for a protocol with non-Gaussian modulation in which the distance $\norm{\vu\rho-\vu\sigma}\leq\epsilon_{\text{prep}}$ and the Gaussian modulation protocol has unconditional security $\epsilon$, the non-Gaussian protocol would then be $(\epsilon+\epsilon_{\text{prep}})$-secure, where $\epsilon_{\text{prep}}$ encompasses a "preparation error" of the thermal state in a Gaussian modulation, derived in \cite{jouguet2012}. The problem with this approach is that in \cite{jouguet2012}, the preparation error refers to the limitation of the modulation device in generating truncated values in the preparation of coherent states and is calculated in the scenario where the running protocol is with Gaussian modulation, such that Bob and Eve expect to observe a thermal state. Therefore, the security factor does not directly apply to the case of discrete modulations, where there is no truncation error in the modulation, and Bob (and Eve) do not expect to observe a thermal state at Alice's laboratory output.



{\appendix[Perturbation Theory in a Nutshell]
In this section, a brief introduction to the theory of perturbation of linear operators will be given, restricted only to the necessary points for the proof of \Cref{prop:spectrum-convergence}. For further details, refer to the textbook 
Consider a Banach space\footnote{A Banach space $X$ is defined as a complete normed vector space, where completeness is related to the existence of limits for every Cauchy sequence $\qty{u_n}$, $u_n,u\in X$. A Hilbert space is then a Banach space whose norm is induced by the inner product.} $X$ and a linear operator $T\in\mathcal{B}(X)$. An eigenvalue of $T$ is defined as a complex number $\lambda$ for which there exists a nonzero element $u$ in the domain of $T$ that satisfies the identity $Tu = \lambda u$, and $u$ is an eigenvector of $T$ associated with the eigenvalue $\lambda$. Similarly, $\lambda$ is an eigenvalue of $T$ if the null space $N$ of the transformation $T-\lambda$ is nontrivial, with $\opname{dim}[N(T-\lambda)] = m$ being the multiplicity of $\lambda$. Assume that $T$ is closed\footnote{An operator $T$ is said to be closed if its domain is a closed set.} in $X$. Then, $T-z$ is closed for every $z\in\mathbb{C}$, and if $T-z$ is invertible,
\begin{equation}
    R(z) = R(z,T) = (T-z)^{-1}\in\mathcal{B}(X),
\end{equation}
is called the resolvent of $T$, and $z$ belongs to the resolvent set of $T$, defined as the complement (with respect to the complex plane) of the spectrum of $T$, i.e., $P(T) = \mathbb{C}\setminus\Lambda(T)$. The resolvent set is a subset of $\mathbb{C}$ for which, given a nonzero element $v\in X$, there exists a solution to the equation
\begin{equation}
    (T-z)u = v.
\end{equation}
Consider the perturbed operator $T(x) = T + xT'$, $x\in\mathbb{C}$. $T(0)$ is called the unperturbed operator, and it is valid to question how the eigenvalue spectrum and eigenvectors of $T(x)$ change as the operator is perturbed. In general, the perturbation will be of the same order of magnitude as $|x|$. As it will be relevant in this work to analyze the perturbation of eigenvalues and eigenvectors according to the norm of the perturbation operator, $\norm{xT'}$, we will use the notation $A = xT'$, and denote the perturbed operator by $T(x) = T + xT' = T+A = T_A$. Represent the resolvent of the perturbed operator $T$ by $A$ as
\begin{equation}
    R_A(z) = (T_A - z)^{-1}.
\end{equation}
It is possible to represent the resolvent as a power series in $x$ (and consequently in $A$), resulting in
\begin{align}
    R_A(z) &= R(z)\sum_{p=0}^\infty[-AR(z)]^p \\
           &= R(z) + R(z)\sum_{p=1}^\infty\qty[-A\cdot R(z)]^p.
\end{align}
where $R(z) = R(z, 0)$, is called the second Neumann series for the resolvent and denotes the effects of the operator $A$ on the resolvent set, which is directly connected to the eigenvalue spectrum. Similarly, the perturbation in the eigenprojectors of $T$ can be observed. Let $\lambda$ be an eigenvalue of $T$ with multiplicity $m$, and let $\Gamma$ be a closed curve in $P(T)$ containing $\lambda$ and no other eigenvalues of $T$. The operator defined by
\begin{align}\label{eq:projector}
    P(x) = P_A &= -\frac{1}{2\pi i}\int_{\Gamma}R_A(z)\dd[2]{z} \\
               &= P_\lambda -\frac{1}{2\pi i}\int_{\Gamma}R(z)\sum_{p=1}^\infty[-AR(z)]^p\dd[2]{z},
\end{align}
is a projector that equals the sum of all eigenprojectors related to the eigenvalues of $T_A$ that lie within $\Gamma$, where $P_\lambda = P(0)$. 

}

\bibliographystyle{IEEEtran}
\bibliography{ref}

\begin{thebibliography}{1}
\providecommand{\url}[1]{#1}
\csname url@samestyle\endcsname
\providecommand{\newblock}{\relax}
\providecommand{\bibinfo}[2]{#2}
\providecommand{\BIBentrySTDinterwordspacing}{\spaceskip=0pt\relax}
\providecommand{\BIBentryALTinterwordstretchfactor}{4}
\providecommand{\BIBentryALTinterwordspacing}{\spaceskip=\fontdimen2\font plus
\BIBentryALTinterwordstretchfactor\fontdimen3\font minus \fontdimen4\font\relax}
\providecommand{\BIBforeignlanguage}[2]{{%
\expandafter\ifx\csname l@#1\endcsname\relax
\typeout{** WARNING: IEEEtran.bst: No hyphenation pattern has been}%
\typeout{** loaded for the language `#1'. Using the pattern for}%
\typeout{** the default language instead.}%
\else
\language=\csname l@#1\endcsname
\fi
#2}}
\providecommand{\BIBdecl}{\relax}
\BIBdecl

\bibitem{holevo2019}
A.~S. Holevo, \emph{Quantum Systems, Channels, Information: A Mathematical Introduction}, 2nd~ed., ser. Texts and Monographs in Theoretical Physics.\hskip 1em plus 0.5em minus 0.4em\relax {Berlin ; Boston}: {De Gruyter}, 2019.

\bibitem{wehrl1976}
A.~Wehrl, ``Three theorems about entropy and convergence of density matrices,'' \emph{Reports on Mathematical Physics}, vol.~10, no.~2, pp. 159--163, Oct. 1976.

\bibitem{leverrier2013}
A.~Leverrier, R.~{Garc{\'i}a-Patr{\'o}n}, R.~Renner, and N.~J. Cerf, ``Security of continuous-variable quantum key distribution against general attacks,'' \emph{Physical Review Letters}, vol. 110, no.~3, p. 030502, Jan. 2013.

\bibitem{kato1995}
T.~Kat{\=o}, \emph{Perturbation Theory for Linear Operators}, ser. Classics in Mathematics.\hskip 1em plus 0.5em minus 0.4em\relax {Berlin}: {Springer}, 1995.

\bibitem{leverrier2015}
A.~Leverrier, ``Composable security proof for continuous-variable quantum key distribution with coherent states,'' \emph{Phys. Rev. Lett.}, vol. 114, no.~7, 2015.

\bibitem{leverrier2017}
------, ``Security of {{Continuous-Variable Quantum Key Distribution}} via a {{Gaussian}} de {{Finetti Reduction}},'' \emph{Phys. Rev. Lett.}, vol. 118, no.~20, 2017.

\bibitem{chen2023}
Z.~Chen, X.~Wang, S.~Yu, Z.~Li, and H.~Guo, ``Continuous-mode quantum key distribution with digital signal processing,'' \emph{npj Quantum Information}, vol.~9, no.~1, p.~28, Mar. 2023.

\bibitem{roumestan2021}
F.~Roumestan, A.~Ghazisaeidi, J.~Renaudier, P.~Brindel, E.~Diamanti, and P.~Grangier, ``Demonstration of probabilistic constellation shaping for continuous variable quantum key distribution,'' in \emph{2021 Optical Fiber Communications Conference and Exhibition ({{OFC}})}, 2021, pp. 1--3.

\bibitem{jouguet2012}
P.~Jouguet, S.~{Kunz-Jacques}, E.~Diamanti, and A.~Leverrier, ``Analysis of imperfections in practical continuous-variable quantum key distribution,'' \emph{Phys. Rev. A}, vol.~86, p. 32309, 2012.

\end{thebibliography}










\newpage

 





\end{document}